\documentclass[aps,prl,superscriptaddress,twocolumn,10pt,nofootinbib,preprintnumbers]{revtex4}

\usepackage{bm}
\usepackage{epsfig}
\usepackage{graphics}
\usepackage{amsmath}
\usepackage{xcolor}

\begin{document}
	
	\title{Collective modes and gapped momentum states in liquid Ga:\\ experiment, theory and simulation}
	
	\author{\firstname{R~M.}~\surname{Khusnutdinoff}}
	\affiliation{Kazan Federal University, 420008 Kazan, Russia}
	
	\author{\firstname{C.}~\surname{Cockrell}} \email{c.j.cockrell@qmul.ac.uk}
	\affiliation{School of Physics and Astronomy, Queen Mary University of London, Mile End Road, London, E1 4NS, UK}
	
	\author{\firstname{O. A.}~\surname{Dicks}}
	\affiliation{School of Physics and Astronomy, Queen Mary University of London, Mile End Road, London, E1 4NS, UK}
	
	\author{\firstname{A. C. S.}~\surname{Jensen}}
	\affiliation{School of Physics and Astronomy, Queen Mary University of London, Mile End Road, London, E1 4NS, UK}
	\affiliation{Department of Chemical Engineering, Imperial College London, London, SW7 2AZ, UK.}
	
	\author{\firstname{M. D.}~\surname{Le}}
	\affiliation{ISIS facility, Rutherford Appleton Laboratory, Chilton, Didcot OX11 0QX Oxfordshire, UK. }
	
	\author{\firstname{L.}~\surname{Wang}}
	\affiliation{School of Physics and Astronomy, Queen Mary University of London, Mile End Road, London, E1 4NS, UK}
	
	\author{\firstname{M. T.}~\surname{Dove}}
	\affiliation{School of Physics and Astronomy, Queen Mary University of London, Mile End Road, London, E1 4NS, UK}
	\affiliation{Department of Physics, School of Sciences, Wuhan University of Technology, 205 Luoshi Road, Hongshan district, Wuhan, Hubei 430070, People's Republic of China}
	\affiliation{College of Computer Science and College of Physical Science \& Technology, Sichuan University, Chengdu 610065, People's Republic of China}
	
	\author{\firstname{A.~V.}~\surname{Mokshin}}
	%\email{anatolii.mokshin@mail.ru}
	\affiliation{Kazan Federal University, 420008 Kazan, Russia}
	
	\author{\firstname{V.~V.}~\surname{Brazhkin}}
	\affiliation{Institute for High Pressure Physics, RAS, 108840 Moscow, Russia}
	
	\author{\firstname{K.}~\surname{Trachenko}}
	\affiliation{School of Physics and Astronomy, Queen Mary University of London, Mile End Road, London, E1 4NS, UK}
	
	
	\begin{abstract}
		Collective excitations in liquids are important for understanding liquid dynamical and thermodynamic properties. Gapped momentum states (GMS) are a notable feature of liquid dynamics predicted to operate in the transverse collective excitations. Here, we combine inelastic neutron scattering experiments, theory, and molecular dynamics modelling to study collective excitations and GMS in liquid Ga in a wide range of temperature and $k$ points. We find that all three lines of enquiry agree for the longitudinal liquid collective dynamics. In the transverse collective dynamics, the experiments agree with theory, modelling as well as earlier x-ray experiments at larger $k$, whereas theory and modelling agree in a wide range of temperature and $k$ points. We observe the emergence and development of the $k$ gap in the transverse dispersion relation which increases with temperature and inverse of relaxation time as predicted theoretically.
	\end{abstract}
	\maketitle
	
	\section{Introduction}
	
	Many important properties of solids are related to collective excitations, phonons \cite{landau}. For example, solid energy and heat capacity are consistently understood on the basis of phonons. For a long time, it remained unclear whether this approach applies to liquids where dynamical disorder and the absence of a fixed lattice seemingly preclude the expansion of vibrational energy around a reference point required to derive phonons as in the solid theory. Calculating the liquid energy as an integral over correlation functions and the interaction potential is highly system dependent due to strong interactions. For this reason, it is believed that liquid thermodynamic properties can not be calculated in general form, contrary to solids and gases \cite{landau}. This singles out the liquid state as the state not amenable to general theoretical treatment. Known more generally as the absence of a small parameter in liquids, this was the long-standing problem in statistical physics research and teaching \cite{granato}.
	
	Phonons in the continuum or hydrodynamic limit of small frequency $\omega$ and wave vector $k$, are sound waves \cite{hydro}. As these phonons make negligible contribution to the density of states, they make only a small contribution to the liquid energy. An important question is whether liquids can support solidlike phonons with large $k$ and $\omega$ approaching the zone boundary as in solids, therefore setting the system energy as in the solid theory. Experiments have ascertained that this is the case for longitudinal phonons \cite{burkel,morkel}, but solidlike transverse phonons have proved to be harder to detect experimentally and understand theoretically. Propagating transverse modes in liquids were first detected in viscous liquids. Later inelastic scattering experiments ascertained the same in low-viscosity liquids such as Na (see Ref. \cite{Trachenko2016} for review). The transverse modes are important because (a) transverse modes have been traditionally viewed as a property of the solid state and are associated with the system having a well-defined shape set by the ability to support shear waves and (b) in solid theory, these modes contribute two thirds of the system energy and heat capacity.
	
	As compared to solids, it took long time to understand the propagation of phonons and in particular solidlike transverse phonons in liquids. Generalised hydrodynamics \cite{boon} was a popular approach which aims to start with the hydrodynamic description where the low-frequency longitudinal sound is the only collective mode in the system and subsequently generalize it to include the non-hydrodynamic solidlike property of nonzero response to shear stress. This results in the prediction that liquids are able to support shear modes as solids do, but only above a finite critical value of the wave vector $k_g$. This is in interesting contrast to the commonly held view that liquids are able to support transverse modes above a certain finite frequency $\omega=1/\tau$ rather than above a finite wave vector \cite{dyre}. In other words, liquids are predicted to have a gap in $k$ space, rather than $\omega$ space.
	
	The prediction of the $k$ gap was indirectly present in the Frenkel theory \cite{frenkel} which preceded generalized hydrodynamics by many decades. Using Maxwell interpolation to describe the viscoelastic liquid response, Frenkel represented viscosity as an operator in the Navier-Stokes equations \cite{pre} but, surprisingly, did not seek the solution of the resulting equation. The solution was shown to give propagating transverse modes with the dispersion relation \cite{Trachenko2016,pre}
	
	\begin{equation}
	\omega=\sqrt{c^2k^2-\frac{1}{4\tau^2}}
	\label{disp}
	\end{equation}
	
	\noindent where $c$ is the transverse speed of sound in the solid and $\tau$ is liquid relaxation time.
	
	In the Maxwell-Frenkel picture, $\tau$ in Eq. (\ref{disp}) is $\eta_s/G_\infty$, where $G_\infty$ is the high-frequency shear modulus and $\eta_s$ is shear viscosity. At the atomistic level, Frenkel theory identified $\tau$ with liquid relaxation time, the average time between molecular jumps around quasiequilibrium positions \cite{frenkel}. This has become an accepted view \cite{dyre}.
	
	Equation (\ref{disp}) predicts that liquids are able to support propagating phonons but only with wave vectors above $k_g$, where
	
	\begin{equation}
	k_g=\frac{1}{2c\tau}
	\label{kg}
	\end{equation}
	
	\noindent which interestingly and importantly differs from solids where $k_g$ is zero or, to be more precise, is a very small number set by the system size.
	Gapped momentum states (GMS) are interesting on their own: for example, Eq. (\ref{disp}) implies an anomalous phase and group velocities as compared to the more common dispersion relation with the energy gap \cite{gapreview}.
	
	It has been realized that in addition to liquids, GMS emerge in a surprising variety of areas \cite{gapreview}, including strongly-coupled plasma, electromagnetic waves, non-linear Sine-Gordon model, relativistic hydrodynamics and holographic models. In some of these areas, GMS are central to the system behavior and are actively studied. In other areas, GMS are not well understood and are often not discussed. A recent review \cite{gapreview} suggests that there is likely to be a common underlying mechanism for GMS in different areas.
	
	As far as liquid theory is concerned, the dispersion relation (\ref{disp}) is important because it can be used to calculate the liquid energy as is done in solids \cite{Trachenko2016,Yang2017}. This calculation predicts a decrease of constant-volume liquid heat capacity with temperature, as is seen experimentally. This decrease is caused by the reduction of the number of phonons propagating above $k_g$ because, according to Eq. (\ref{kg}), the range of $k$ points where phonons propagate shrinks with temperature ($\tau$ decreases with temperature) \cite{Yang2017}. In other words, the phase space of phonons in liquids decreases with temperature, in notable contrast to solids.
	
	Although GMS have been seen in simulations of both liquids and supercritical fluids below the Frenkel line
	\cite{MokshinJETP2015,Yang2017,Bolmatov2015}, there is no experimental evidence for it in liquids. As recently reviewed \cite{gapreview}, the only experimental evidence for GMS comes from imaging particles in strongly-coupled plasma \cite{plasma}. This has set one of the motivations behind this work.
	
	Ga is also a substance with a number of unusual properties. This trivalent element is one of the few metallic-like systems that does not crystallize into any simple structure. It shows an extremely rich polymorphism that includes a stable, low-pressure phase, $\alpha$-Ga (orthorhombic structure with eight atoms per unit cell), and two other phases which are stable at high pressure: Ga-II (a body centered cubic phase with twelve atoms per unit cell) and Ga-III (tetragonal phase). Furthermore, the phase diagram of gallium contains a number of metastable phases known as $\beta$, $\gamma$, $\delta$, and $\epsilon$ with melting points of $256.8$, $237.6$, $253.8$, and $244.6$~K all well below that of $302.93$~K for $\alpha$-Ga. The wide temperature range of the existence of the liquid phase ($303-2500$~K) and the low pressure of saturated vapors at $T<1400$~K makes gallium a promising coolant with outstanding thermohydraulic properties. These properties, together with low melting point, make Ga an interesting and convenient system to study experimentally. Ga has received substantial experimental and computational interest in the past. The question of whether gallium can sport acoustic collective modes \cite{Bermejo1994,Bermejo1997} has been answered, with longitudinal excitations having been seen in inelastic neutron and x-ray scattering experiments at a wide range of wave vectors and temperatures \cite{Bermejo1997,Scopigno2002,Bove2005}. High frequency excitations resembling solidlike optical modes have also been detected \cite{Bermejo2005}, setting Ga apart from ``simple" molten metals like Na or Rb. The possibility that these excitations were diffusive/incoherent contributions has been ruled out \cite{Patty2011}. However, no explanations in terms of coherent excitations can account for the large observed intensity, leaving the phenomenon an open question. Ga is therefore a remarkable and interesting substance for its thermodynamic, structural, and dynamical properties.
	
	In this paper, we combine inelastic neutron scattering experiments, modelling, and theory to study phonons in liquid Ga in a wide range of temperatures, frequencies, and wave numbers. We find that all three methods agree for the longitudinal phonons. For transverse phonon branch, we find agreement between (a) our modelling and theoretical results with inelastic neutron and x-ray scattering at higher $k$ and (b) theory and modelling data in the entire range of $k$ points showing the $k$ gap. We observe the increase of the $k$ gap with temperature and inverse of relaxation time, in agreement with Maxwell-Frenkel theory. This agreement, together with the agreement of all three lines of enquiry for the longitudinal mode, builds up the body of evidence for GMS. Our results serve as a stimulus for future inelastic neutron scattering experiments investigating transverse modes in liquids and its evolution in terms of gapped momentum states.
	
	\section{Inelastic Neutron Scattering Experiment}
	%~~~~~~~~~~~~~~~~~~~~~~~~figure~~~~~~~~~~~~~~~~~~~~~~~~~~~~~~~~~~~~~~~~~~~~
	\begin{figure*}
		\begin{center}
			\includegraphics[keepaspectratio,width=\linewidth]{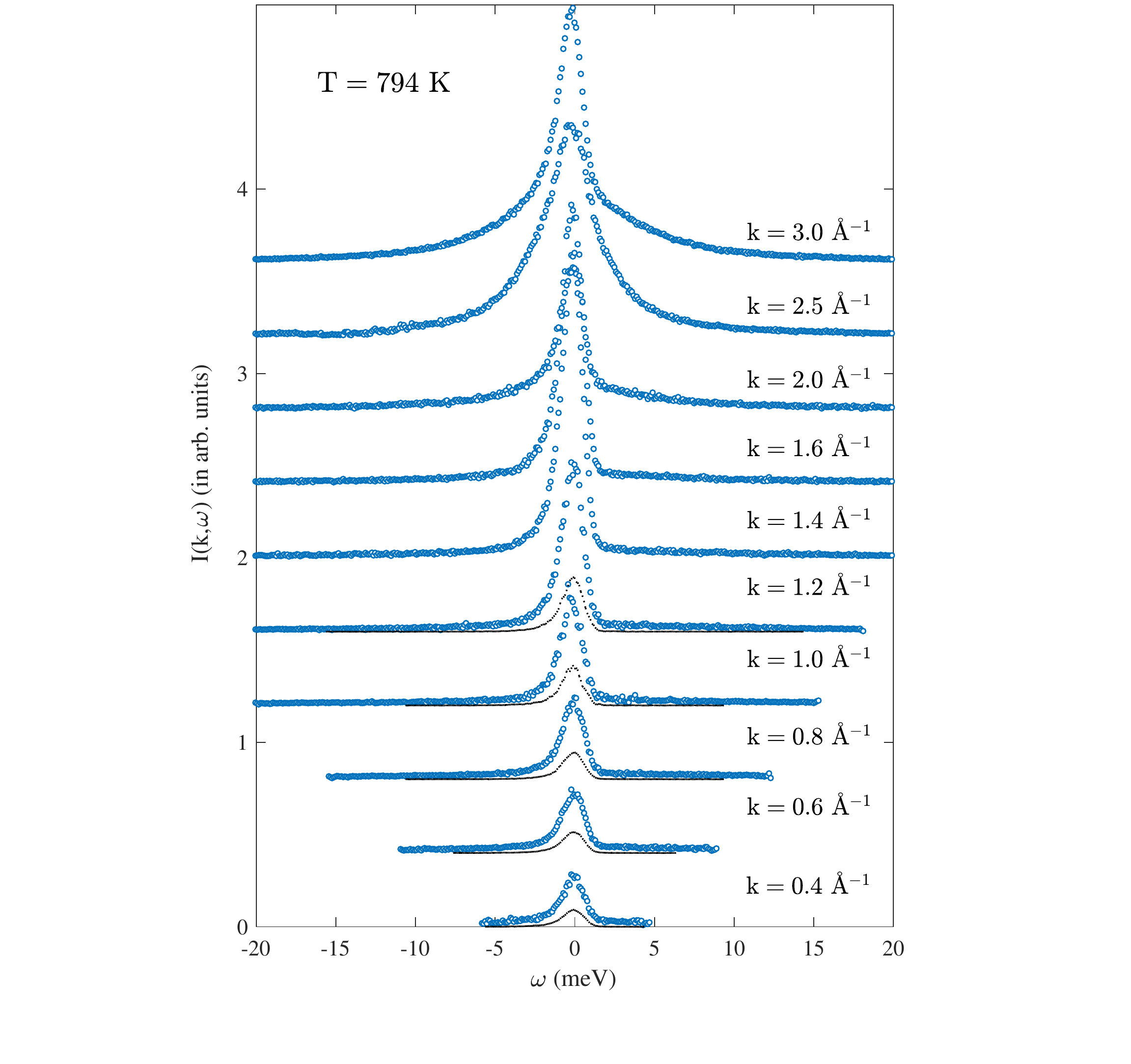} %[height=9cm, angle=0]{fig01.eps}
			\caption{Intensity neutron scattering $I(k,\omega)$ in molten gallium at $T=794$~K at the fixed wave numbers. The experiment was performed on an instrument (MARI at ISIS) where the minimum scattering angle is 3$^o$ and the monochromatic (pulsed) beam had an energy $40$~meV. Dotted line represents the spectrometer resolution function.}
			\label{Fig_Ikw}
		\end{center}
	\end{figure*}
	%~~~~~~~~~~~~~~~~~~~~~~~~figure~~~~~~~~~~~~~~~~~~~~~~~~~~~~~~~~~~~~~~~~~~~
	Seven molten gallium samples were prepared in vacuum-sealed quartz glass cylinders with 2.5 mm radius. The neutron scattering experiment (INS) was performed at the MARI chopper spectrometer at ISIS with incident neutron energies of 7 meV and 40 meV \cite{MDove}. The energy resolution was $\Delta E = 0.3$ to $2$ meV, for incident energies 7 to 40 meV at wave numbers 0.4 to 1.5 $\rm{\AA}^{-1}$. We collected our data at six temperature points in a wide temperature range: 313, 400, 500, 600, 700 and 794~K. The INS spectra of liquid gallium $I(k,\omega)$ at the temperature $794$~K as a function of energy transfer for several constant $k$ values is shown in Fig.~\ref{Fig_Ikw}. Here, dotted line represents the spectrometer resolution function.
	
	Initially, raw experimental data were processed as follows. All data on the intensity neutron scattering in liquid gallium were centered at zero frequency. To improve the quality of the INS spectra at fixed wave numbers $k$, we averaged the spectra over the wave number range $k\in[k-\Delta k, k+\Delta k]$, where $\Delta k$ was about $10\%$ of $k$.
	
	The experimental spectra were analyzed in terms of a model that accounts for coherent and incoherent contributions to the intensity neutron scattering is given by~\cite{Scopigno_RMP}:
	\begin{equation}
	I(k,\omega)=E(k)\int R(k,\omega-\omega')S_q(k,\omega')d\omega'+B(k,\omega),
	\label{Eq_INSscatter}
	\end{equation}
	
	\begin{align}
	\begin{split}
	S_q(k,\omega)= & \frac{\hbar\beta\omega}{1-e^{-\hbar\beta\omega}}\bigg[\frac{\sigma_\mathrm{coh}}{\sigma_\mathrm{coh}+\sigma_\mathrm{incoh}}
	S(k,\omega) \\
	& +\frac{\sigma_\mathrm{incoh}}{\sigma_\mathrm{coh}+\sigma_\mathrm{incoh}}S_s(k,\omega)\bigg].
	\label{Eq_INSsqw}
	\end{split}
	\end{align}
	Here, $E(k)$ represents a scaling parameter; $S_q(k,\omega)$ and $S_s(k,\omega)$ are the quantum dynamic and self-dynamic structure factors, respectively;  the term $B(k,\omega)$ corresponds to the background; $R(k,\omega)$ is the spectrometer resolution; $\beta=1/(k_BT)$ is the reciprocal temperature; $\sigma_\mathrm{coh}$ and $\sigma_\mathrm{incoh}$ are the cross sections of coherent and incoherent scattering, respectively [for Ga $\sigma_\mathrm{incoh}/(\sigma_\mathrm{incoh}+\sigma_\mathrm{coh})=0.0658$] \cite{Funel_1998}.
	
	The measured signal from a vanadium standard sample at different scattering angles was used as an experimental estimate the spectrometer resolution $R(k,\omega)$,  which we then fitted to the function
	\begin{eqnarray}
	R(k,\omega) =\frac{1}{\sqrt{2\pi\omega_0(k)^2}}\exp\bigg(-\frac{\omega^2}{2\omega_0^2(k)}\bigg), \label{Eq_Resol}
	\end{eqnarray}
	satisfying the normalization condition
	\begin{eqnarray}
	\int_{-\infty}^{\infty}R(k,\omega)d\omega=1. \nonumber
	\end{eqnarray}
	Here, $\omega_0(k)$ corresponds to the standard Gaussian deviation.
	
\begin{figure*}
		\begin{center}
			\includegraphics[keepaspectratio,width=\linewidth]{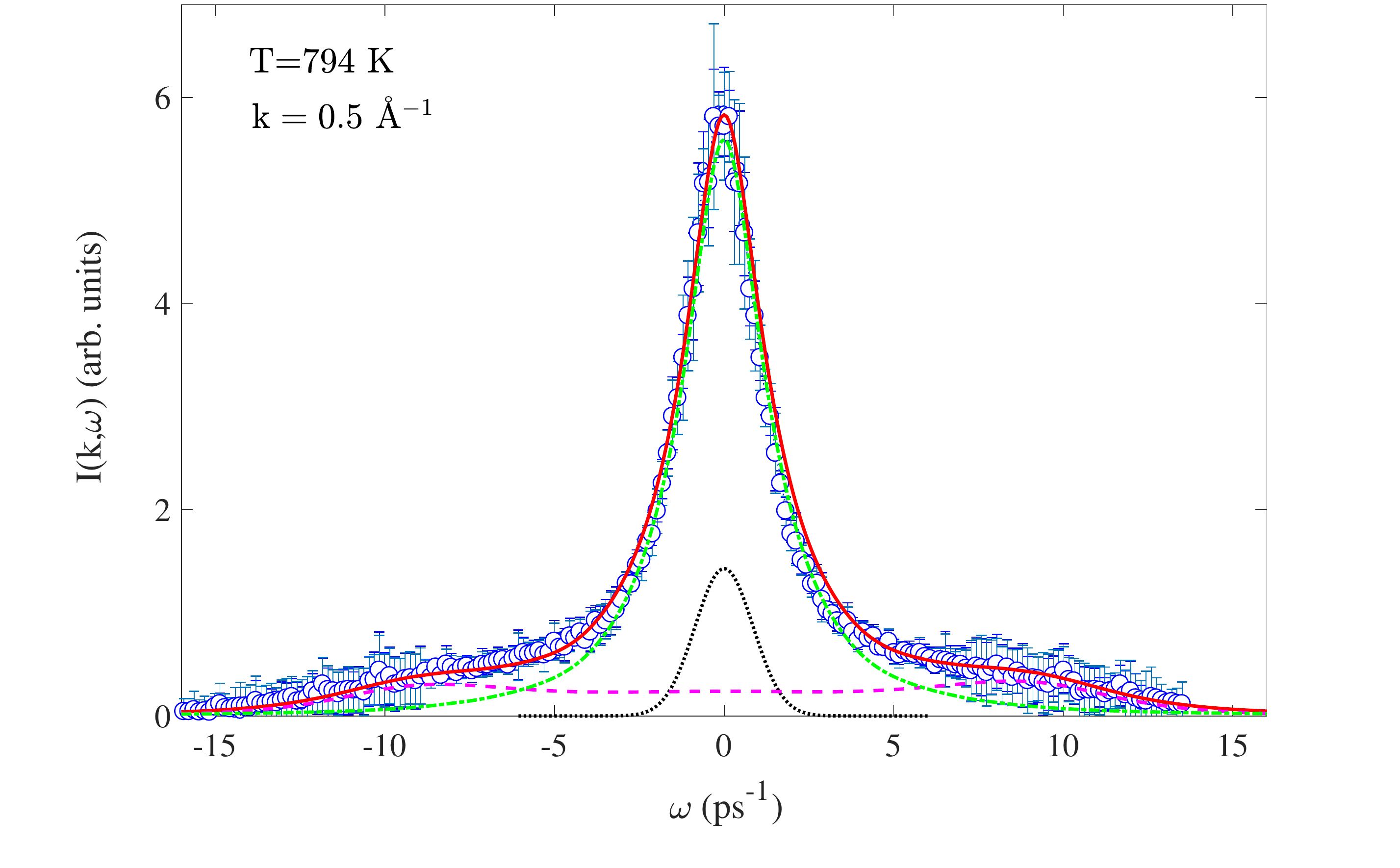} %[height=11cm, angle=0]{fig02.pdf}
			\caption{Intensity neutron scattering $I(k,\omega)$ in molten gallium. Dashed magenta line -- the dynamic structure factor calculated from molecular dynamics simulation data; green line -- the resulting function containing partial components associated with various types of neutron scattering on matter: multiple, quasielastic coherent and quasielastic incoherent scattering; dotted black line -- the spectrometer resolution function; solid red line -- ``theoretical'' intensity of neutron scattering in liquid gallium, obtained taking into account the experimental resolution function.}
			\label{Fig_Ikw2}
		\end{center}
	\end{figure*}

	The contribution of gallium's moderate neutron absorption cross section was calculated and compensated for by comparison with the background high-$k$ peaks from the niobium furnace. The self-dynamic structure factor (incoherent scattering) was modeled using the simple hydrodynamic formula \cite{Berrod2018}:
	\begin{eqnarray}
	S_s(k,\omega) =\frac{1}{\pi}\exp\bigg(\frac{D k^2}{\omega(k)^2 + (D k^2)^2}\bigg), \label{Eq_Sskw}
	\end{eqnarray}
	\noindent where $D$ is the self-diffusion coefficient obtained from the molecular dynamics simulations. As we have shown in Ref.~\cite{MokshinJETP2015}, the results of MD simulations correctly predict the temperature dependence of the self-diffusion coefficient in liquid gallium.
	Then, we have attempted to reproduce neutron scattering intensity spectra with a dynamic structure factor (coherent scattering) obtained from molecular dynamics simulations. Despite the fact that the calculated dynamic structure factor correctly reproduces the collective dynamics of liquid gallium and gives the correct values of the static structure factor
	\begin{eqnarray}
	S(k)=\int_{-\infty}^{\infty}S(k,\omega)d\omega \nonumber
	\end{eqnarray}
	we were unable to correctly reproduce all the features of the $I(k,\omega)$ spectra.
	As can be seen from Fig.~\ref{Fig_Ikw}, the collective acoustic branches are strongly suppressed by the huge central contribution.
	In order to correctly reproduce all features of the INS spectra, the central elastic line was fitted with an additional Lorentzian function.
	This contribution, as was noted in Ref.~\cite{Blagoveshchenskii2014} can be responsible for multiple, quasielastic coherent and quasielastic incoherent scattering.  The intensity neutron scattering $I(k,\omega)$ in molten gallium at the temperature $T=794$~K and its partial components are presented in Fig.~\ref{Fig_Ikw2}.
	
	When calculating current autocorrelation spectra, we used the experimental scattering intensely directly. In order to extract the transverse dispersion curves from the current INS data we used fitting procedure with the two-oscillator model \cite{Kryuchkov2019} and the damped harmonic oscillator model \cite{Monaco2010}.

	%~~~~~~~~~~~~~~~~~~~~~~~~figure~~~~~~~~~~~~~~~~~~~~~~~~~~~~~~~~~~~~~~~~~~~~
	%~~~~~~~~~~~~~~~~~~~~~~~~figure~~~~~~~~~~~~~~~~~~~~~~~~~~~~~~~~~~~~~~~~~~~

	\section{Molecular Dynamics Simulations}
	
	Molecular dynamics (MD) simulations have been performed in the $NPT$ ensemble for the system consisting of $32000$ particles interacting via the EAM potential \cite{Belashchenko_2012} at the temperatures $T=[313,\; 400,\; 500,\; 600,\; 700,\; 794]$~K and pressure of about $1.0$~bar.
	%~~~~~~~~~~~~~~~~~~~~~~~~figure~~~~~~~~~~~~~~~~~~~~~~~~~~~~~~~~~~~~~~~~~~~~
	\begin{figure*}
		\begin{center}
			\includegraphics[keepaspectratio,width=\linewidth]{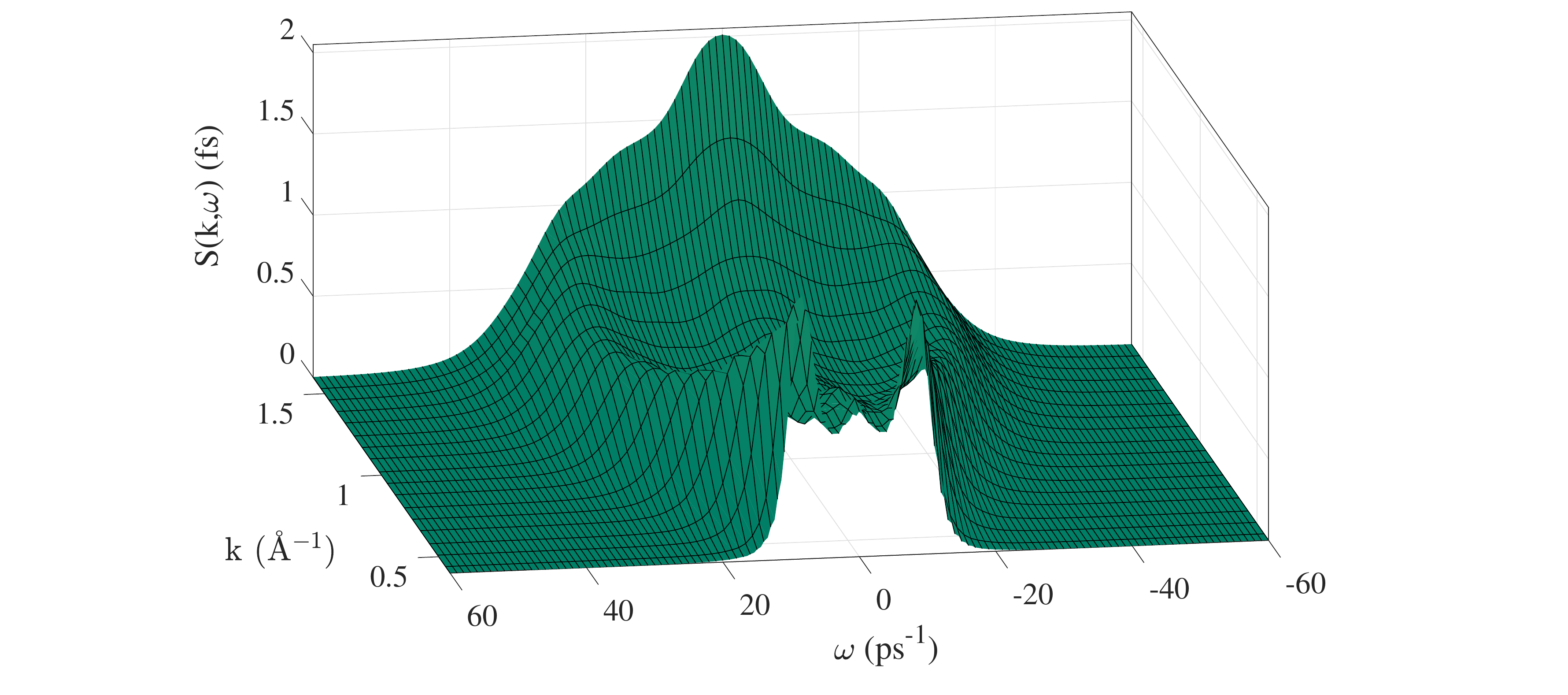} %[height=8cm, angle=0]
			\caption{Dynamic structure factor of liquid gallium calculated from molecular dynamics simulations.}
			\label{Fig_Skw}
		\end{center}
	\end{figure*}
	%~~~~~~~~~~~~~~~~~~~~~~~~figure~~~~~~~~~~~~~~~~~~~~~~~~~~~~~~~~~~~~~~~~~~~
	The dynamic structure factor can be computed on the basis of MD results as
	\begin{equation}
	S(k,\omega)=\frac{S(k)}{2\pi}\textrm{Re}\int_{-\infty}^{\infty}\frac{\big\langle \delta\rho^*(k,0)\delta\rho(k,t) \big\rangle}{\big\langle |\delta\rho(k,0)|^2 \big\rangle} e^{i\omega t}dt,
	\end{equation}
	and its plotted in Fig.~\ref{Fig_Skw}. We also calculate the spectral densities
	\begin{equation}
	\widetilde{C}_{\alpha}(k,\omega)=\frac{1}{2\pi}\int_{-\infty}^{\infty}C_{\alpha}(k,t)e^{i\omega t}dt, ~~~~~~ \alpha\in \{L,T\}
	\end{equation}
	of the time correlation functions (TCF) of the longitudinal current $C_{L}(k,t)$ and the transverse current $C_{T}(k,t)$,
	\begin{equation}
	C_{L}(k,t)=\frac{1}{N}\bigg\langle j_z(\textbf{k},t)j_z(-\textbf{k},0) \bigg\rangle,
	\end{equation}
	\begin{equation}
	\label{Eq_CT}
	C_{T}(k,t)=\frac{1}{2N}\bigg\langle j_x(\textbf{k},t)j_x(-\textbf{k},0) +j_y(\textbf{k},t)j_y(-\textbf{k},0) \bigg\rangle.
	\end{equation}
	
	Here, $\delta\rho(k,t)$ is the density fluctuations, $\textbf{j}(\textbf{k},t)=\sum_l^N\textbf{v}_l(t)e^{-i\textbf{k}\cdot\textbf{r}_l(t)}$ is the velocity current and the wave-vector $\textbf{k}$ is directed along the $z$ axis. Dispersion curves of longitudinal and transverse excitations are obtained from the location of maxima of the spectral densities $\widetilde{C}_L(k,\omega)$ and $\widetilde{C}_T(k,\omega)$ at various values of $k$. The longitudinal current spectra are related to the dynamic structure factor by the equation
	\begin{eqnarray}
	\widetilde{C}_L(k,\omega)=\frac{\omega^2}{k^2}S(k,\omega).
	\label{Eq_CL_Skw}
	\end{eqnarray}
	
	\section{Collective Excitations and Dispersion Relation}
	Using the memory function formalism \cite{Hansen/McDonald,Balucani_1994} in the framework of the theory of self-consistent relaxation \cite{Mokshin_JPCM,MokshinJPCM2018,YulmetyevPRE2001,YulmetyevJETP2002,YulmetyevJPCM2003}, the dynamic structure factor $S(k,\omega)$ is defined as follows
	\begin{eqnarray}
	\label{eq: Basic_a}   
	S(k, \omega)&=& \frac{S(k)}{2 \pi}\frac{\Delta_{1}^2(k) \Delta_{2}^2(k)\Delta_{3}^2(k)}{\Delta_4^2(k)-\Delta_3^2(k)}\times\\ 
	&&\times\frac{[4\Delta_{4}^2(k)- \omega^{2}]^{1/2}}{\omega^6 +\mathcal{A}_1(k)
		\omega^4 +\mathcal{A}_2(k) \omega^2 +\mathcal{A}_3(k)}\nonumber
	\end{eqnarray}
	with
	\begin{eqnarray} \label{eq: coeff_A}
	\mathcal{A}_1(k)&=&
	\frac{\Delta_3^4(k)-\Delta_2^2(k)[2\Delta_4^2(k)-\Delta_3^2(k)]}{\Delta_4^2(k)-\Delta_3^2(k)}
	-2\Delta_1^2(k),\nonumber\\
	\mathcal{A}_2(k)&=&\frac{1}{\Delta_4^2(k)-\Delta_3^2(k)}
	\Bigg[\Delta_2^4(k)\Delta_4^2(k)-2\Delta_1^2(k)\Delta_3^4(k)+\nonumber\\
	&&+\Delta_1^2(k)\Delta_2^2(k)
	[2\Delta_4^2(k)-\Delta_3^2(k)]\Bigg]
	+ \Delta_1^4(k),\nonumber\\
	\mathcal{A}_3(k)&=&\frac{\Delta_1^4(k)\Delta_3^4(k)}{\Delta_4^2(k)-\Delta_3^2(k)}.
	\nonumber
	\end{eqnarray}
	Here, $\Delta_n^2(k)$ (where $n=1,2,3,4$) are the frequency relaxation parameters of the dynamic structure factor. The parameters $\Delta_n^2(k)$ are defined by the following expressions
	\begin{equation} \label{eq: frequency_parameter}
	\Delta_{n}^2(k) = \frac{\langle |W_{n}(k)|^2 \rangle}{\langle |W_{n-1}(k)|^2 \rangle},
	\nonumber
	\end{equation}
	where $W_{n}(k)$ are the dynamical variables, which is associated with the particle density, longitudinal
	momentum component, energy density, and so on; the brackets $\langle \ldots \rangle $ denote the ensemble average. This means that the coupling between the density and energy (thermal fluctuations) are also included within the theoretical scheme ~\cite{MokshinJPCM2018}. The calculation of the frequency relaxation parameters was performed on the basis of MD data using the numerical algorithm presented in Ref.~\cite{MokshinJPCM2018}.
	
	Then substituting the expression for the dynamic structure factor $S(k,\omega)$ into equation~(\ref{Eq_CL_Skw}), we obtain
	\begin{eqnarray}\label{Eq_CL}
	\widetilde{C}_L(k,\omega)&=& \frac{\omega^2S(k)}{2 \pi k^2}\frac{\Delta_{1}^2(k) \Delta_{2}^2(k)\Delta_{3}^2(k)}{\Delta_4^2(k)-\Delta_3^2(k)}\times\\
	&&\times\frac{[4\Delta_{4}^2(k)- \omega^{2}]^{1/2}}{\omega^6 +\mathcal{A}_1(k)
		\omega^4 +\mathcal{A}_2(k) \omega^2 +\mathcal{A}_3(k)}.\nonumber
	\end{eqnarray}
	In the $\omega\to 0$ limit and as the consequence of continuity equation, we have
	\begin{eqnarray}
	\lim_{\omega\to 0}\widetilde{C}_L(k,\omega)&=& 0.
	\end{eqnarray}
	
	Within the framework of this approach, the dispersion law is determined by the expression~\cite{MokshinJPCM2018}:
	\begin{eqnarray} \label{Eq_Disp}
	\omega_c^{(L)}(k) = \sqrt{\frac{-\mathcal{A}_1(k)+\sqrt{\mathcal{A}_1(k)^2-3\mathcal{A}_2(k)}}{3}}.
	\end{eqnarray}
	
\begin{figure*}
		\begin{center}
			\includegraphics[keepaspectratio,width=\linewidth]{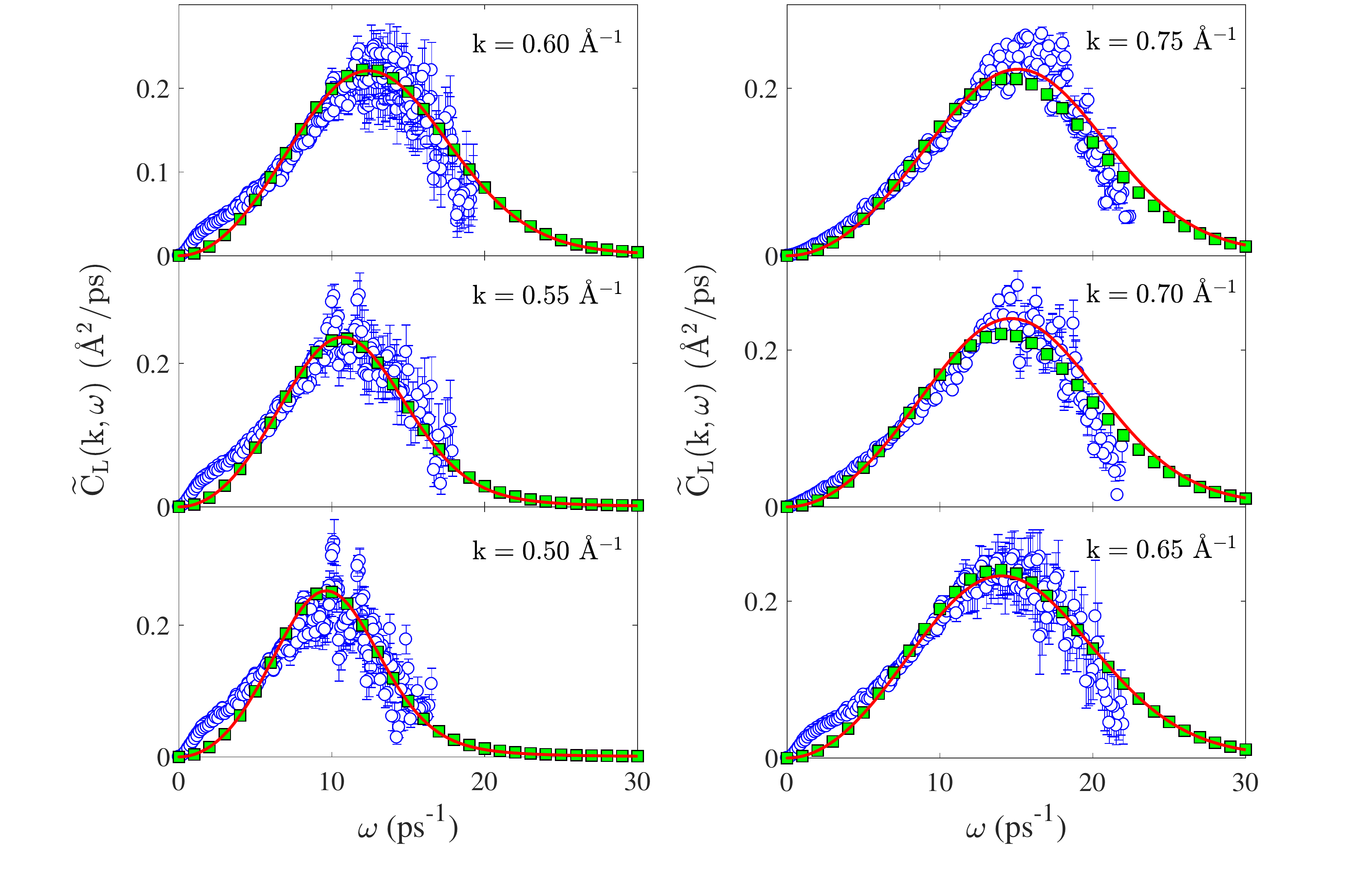} %[height=11cm, angle=0]{fig04.eps}
			\caption{Longitudinal current spectra $\widetilde{C}_L(k,\omega)$ calculated by equation (\ref{Eq_CL_Skw}) for liquid gallium at the temperature $T=794$~K: circles -- experimental results from inelastic neutron scattering; squares -- the molecular dynamics simulation results; solid lines -- theoretical results [Eq.~(\ref{Eq_CL})].}
			\label{Fig_CL}
		\end{center}
	\end{figure*}

	Comparison of the theoretical results of the longitudinal current spectra $\widetilde{C}_L(k,\omega)$ of liquid gallium at $T=794$~K according to equation~(\ref{Eq_CL}) with INS and MD results is presented in Fig.~\ref{Fig_CL}, demonstrating good qualitative and quantitative agreement between all three methods. 
	
\begin{figure*}
		\begin{center}
			\includegraphics[height=8.5cm, angle=0]{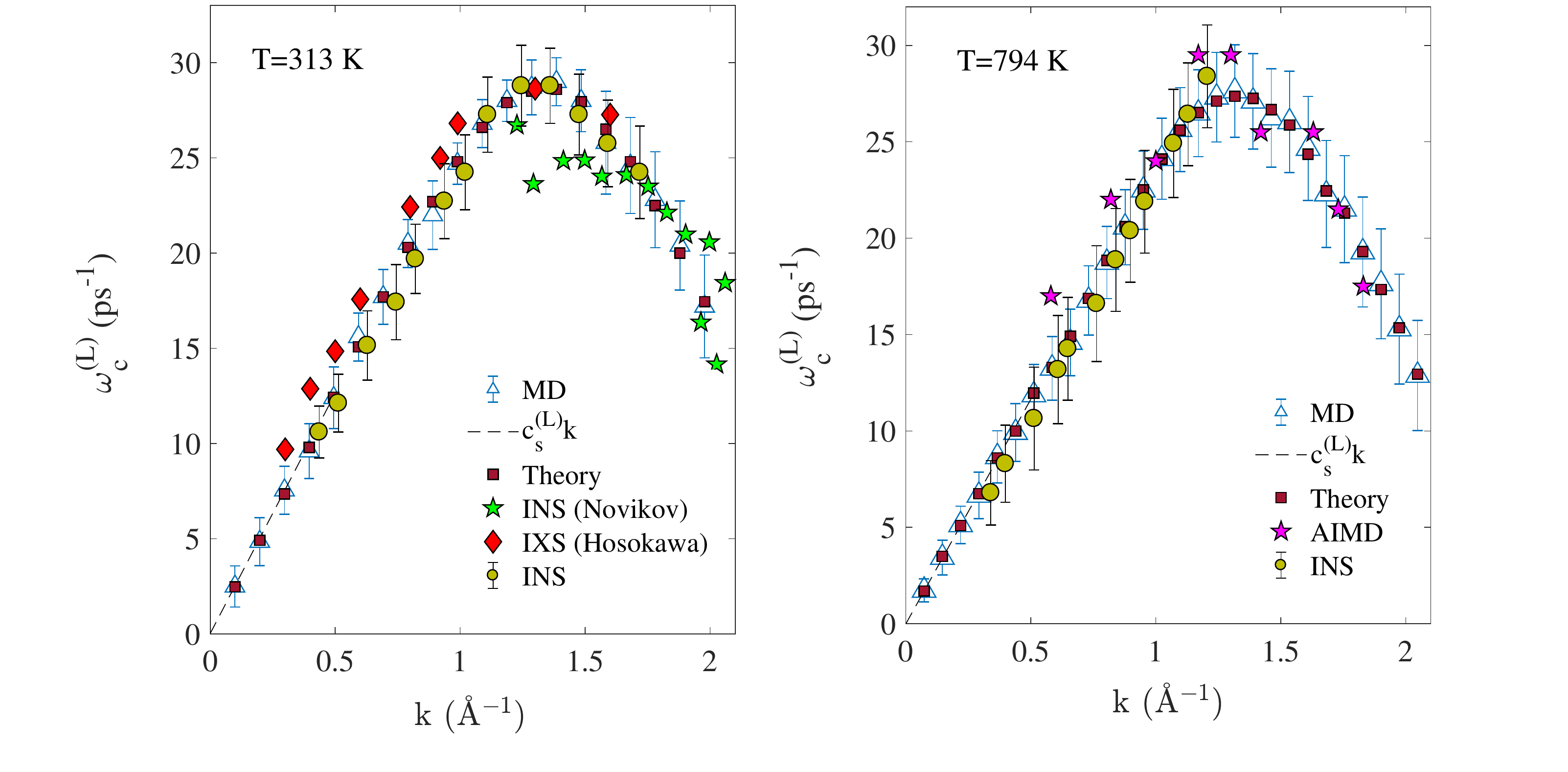} %[height=7.5cm, angle=0]{fig05.eps}
			\caption{Dispersion of the high-frequency peak $\omega_c(k)$ of the longitudinal current spectra $\widetilde{C}_L(k,\omega)$ for liquid gallium at the temperatures LEFT: $T=313$~K: in part reproduced from \cite{MokshinJETP2015}, here circles -- our experimental INS results; RIGHT: $T=794$~K: circles -- the experimental INS results; triangles -- the molecular dynamics simulation results; squares -- theoretical results~(\ref{Eq_Disp}); stars -- AIMD data at the $T=702$~K \cite{Holender1995}. Dashed lines represent the extrapolated hydrodynamic result $\omega_c(k)=c_sk$, where $c_s$ is the sound velocity.}
			\label{Fig_Disp}
		\end{center}
	\end{figure*}

	The dispersion law $\omega_c^{(L)}(k)$ for the liquid gallium at the temperatures $T = 313~K$ and $T=794$~K is given in Fig.~\ref{Fig_Disp}. Here, results of molecular dynamics simulations, INS experiments, and inelastic x-ray scattering (IXS) experiments (reproduced from Ref.~\cite{MokshinJETP2015}) are compared to theoretical results (\ref{Eq_Disp}). As seen, the calculated theoretical and MD dispersion curves are in a good agreement with the INS and IXS data.

	\section{Transverse Dispersion Law and gapped momentum states}
	
\begin{figure*}
		\begin{center}
			\includegraphics[height=11.0cm, angle=0]{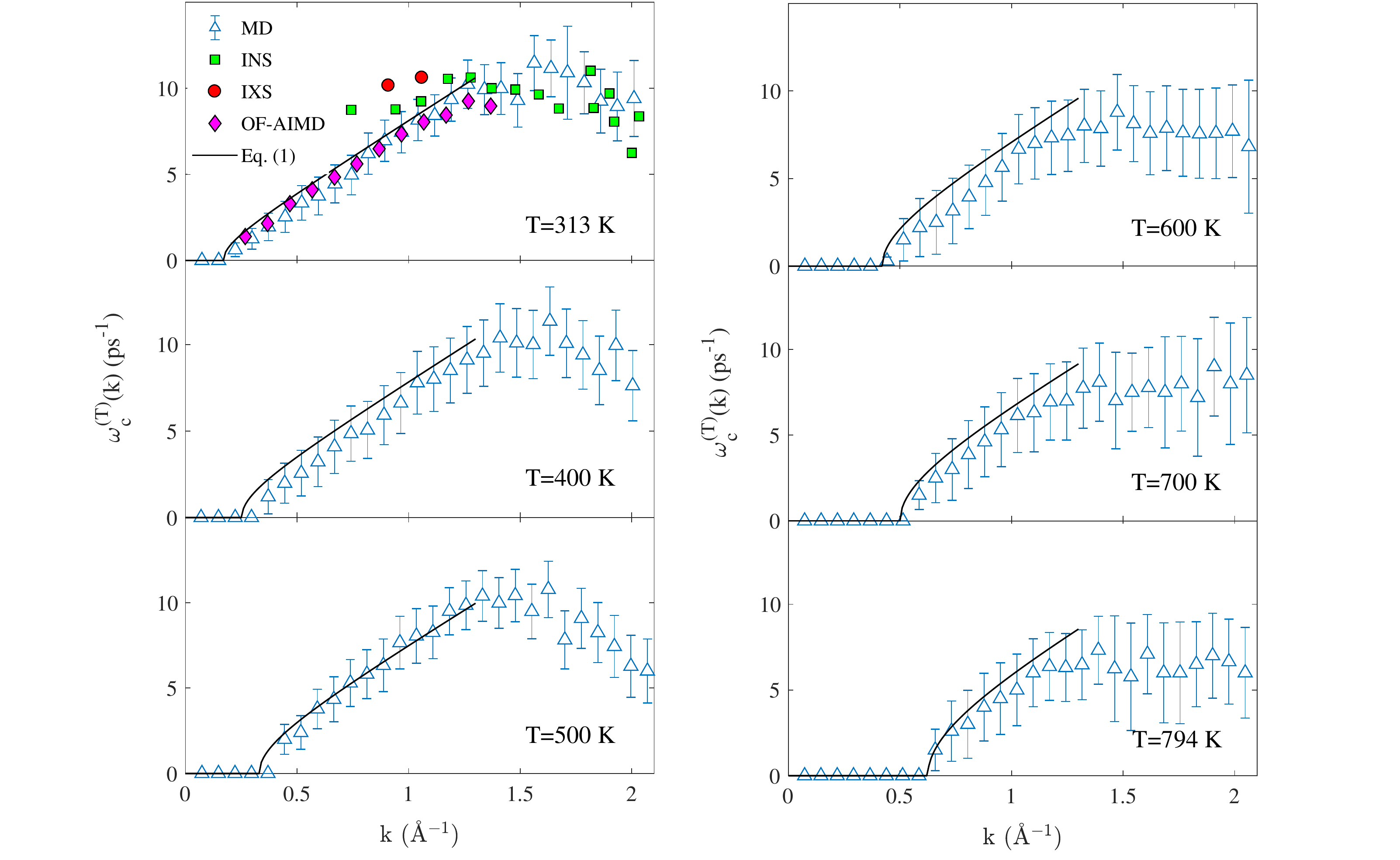} %[height=7.5cm, angle=0]{fig05.eps}
			\caption{The laws of dispersion of the transverse acousticlike excitations of a gallium melt at various temperatures: Open triangles show the results of molecular dynamics simulation; solid squares show the experimental data on inelastic neutron scattering \cite{MokshinJETP2015}; solid circles show the experimental data on inelastic x-ray scattering \cite{Hosokawa2009}; solid diamonds are the results of the OF-AIMD simulation \cite{Hosokawa2009}; solid lines -- theoretical results [Eq.~\ref{disp}].}
			\label{Fig_TransDisp}
		\end{center}
	\end{figure*}

	The transverse current TCFs [Eq. (\ref{Eq_CT})] were calculated from the MD simulations data at different temperatures. The dispersion law of the transverse collective mode calculated from the spectral densities of these TCFs at different temperatures is presented in Fig.~\ref{Fig_TransDisp} alongside experimental dispersion curves from IXS reported previously \cite{Hosokawa2009} as well as earlier \textit{ab initio} modelling results \cite{MokshinJETP2015}.
	
	\begin{figure*}
		\begin{center}
			\includegraphics[height=11.5cm, angle=0]{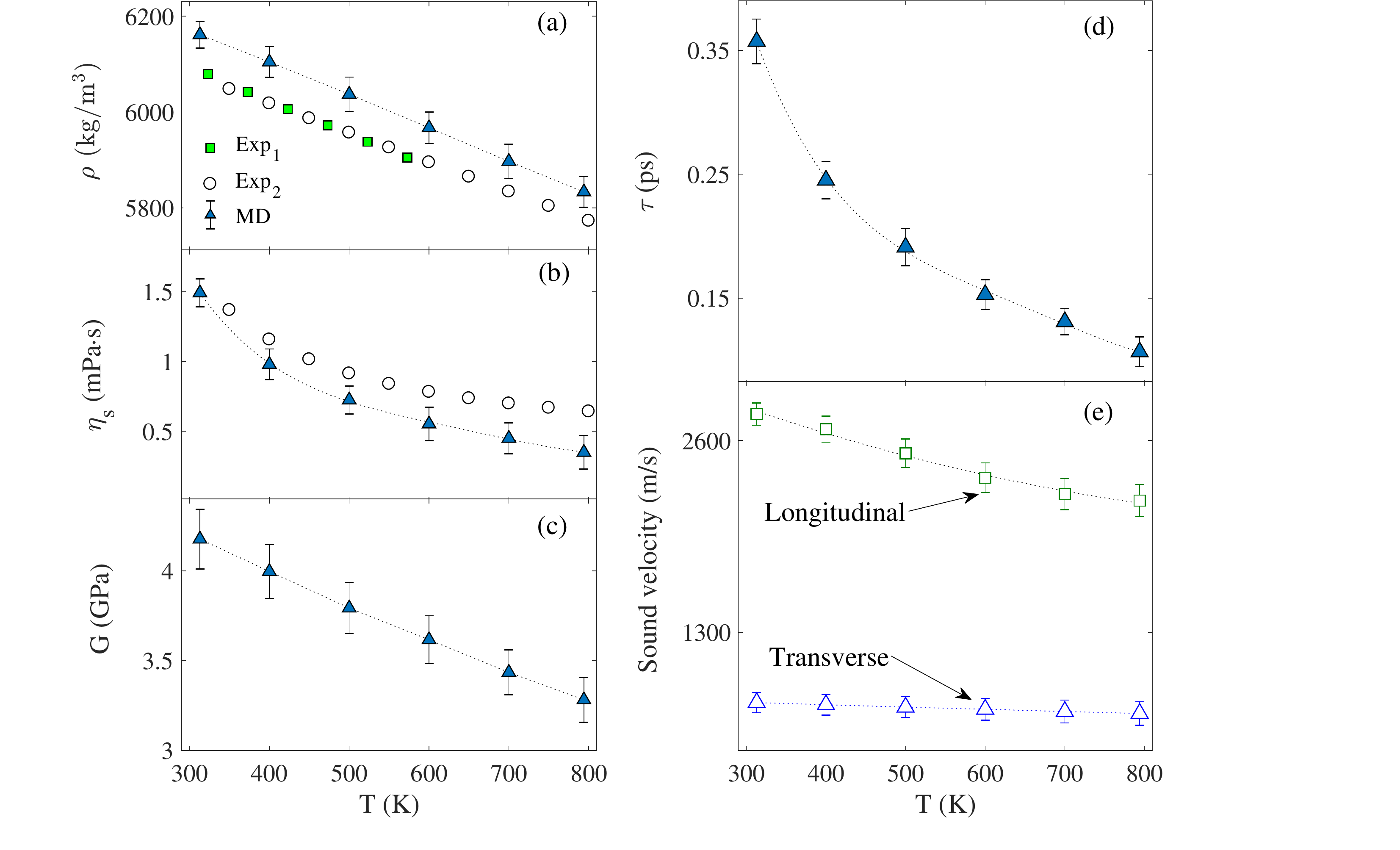} %[height=7.5cm, angle=0]{fig05.eps}
			\caption{Temperature dependencies of mass density $\rho(T)$ of the system (a), shear viscosity $\eta_s(T)$ (b), shear modulus $G$ (c)  and relaxation time $\tau$ (c): Solid triangles -- molecular dynamics simulation results; solid squares and solid circles correspond to the experimental data \cite{Ayrinhac2015} and \cite{Assael2012}, respectively. (e), Temperature dependencies of the sound velocity in liquid gallium: open squares -- longitudinal waves;
				open triangles -- transverse waves.}
			\label{Fig_DensVisc}
		\end{center}
	\end{figure*}
	
	At lowest temperature, we observe an agreement between the calculated dispersion curves and previous experimental results at large $k$. We note that INS and IXS experiments probe density fluctuations and therefore can only detect the presence of transverse modes if those modes are not purely transverse. A liquid's lack of translational invariance precludes purely transverse modes at lower values of $k$ than would be possible for solids \cite{Ruocco1999}. It is by this mechanism that previous INS and IXS experiments have detected transverse modes in Ga, other molten metals, and liquid water at larger $k$ \cite{MokshinJETP2015,Hosokawa2009,Monaco2010,Hosokawa2013,Hosokawa2015, Cimatoribus2010}, usually detectable by a shoulder on or broadening of the elastic peak in the dynamic structure factor. We have attempted to extract the transverse dispersion curves from the current INS data by fitting to a two-oscillator model \cite{Kryuchkov2019} and damped harmonic oscillator model \cite{Monaco2010}, however the resolution and signal strength of the experimental data were lacking and the additional degrees of freedom available in the two-oscillator model resulted in overfitting.
	
	We now focus on modelling and theoretical results for the transverse dispersion curve. First, we calculate the parameters in the predicted transverse dispersion curve (\ref{disp}). We calculate $\tau$ as $\eta_s/G_\infty$, and $\eta_s$ as
	
	\begin{equation}
	\eta_s=\frac{V}{k_BT}\int_0^{\infty}\langle \sigma_{\alpha\beta}(t)\sigma_{\alpha\beta}(0)\rangle dt,
	\end{equation}
	
	\noindent where angle brackets mean averaging over time and ensemble of particles, $k_B$ is the Boltzmann constant, $V$ is the volume, $\sigma_{\alpha\beta}$ are the nondiagonal components of the stress tensor.
	
	We subsequently calculate $c$ as $\sqrt\frac{G_\infty}{\rho}$, where the infinite-frequency shear modulus is evaluated as
	\begin{equation}
	G_\infty=\frac{V}{k_BT}\langle |\sigma_{\alpha\beta}(0)|^2\rangle.
	\end{equation}
	
	Figure \ref{Fig_DensVisc} shows the calculated values of density, viscosity, $G$, speed of sound and relaxation time as a function of temperature. As seen from the figure, the simulation results for the density and viscosity of liquid gallium are in qualitative agreement with the experimental data, correctly predicting the general trend with increasing temperature.
	
	We now address the central point of this paper of how well the dispersion relation for GMS describes liquid dynamics. Our results enable us to study this point quantitatively and in detail.
	
	We use three ways to study this. First, we plot Eq. (\ref{disp}) in Fig. \ref{Fig_TransDisp} using the calculated $\tau$ and $c$. We observe a good agreement with (a) dispersion relation obtained from modelling in the wide rage of $k$ points including the $k$ gap at all temperatures and (b) experimental transverse points at the lowest temperature and higher $k$.
	
\begin{figure*}
		\begin{center}
			\includegraphics[keepaspectratio,width=\linewidth]{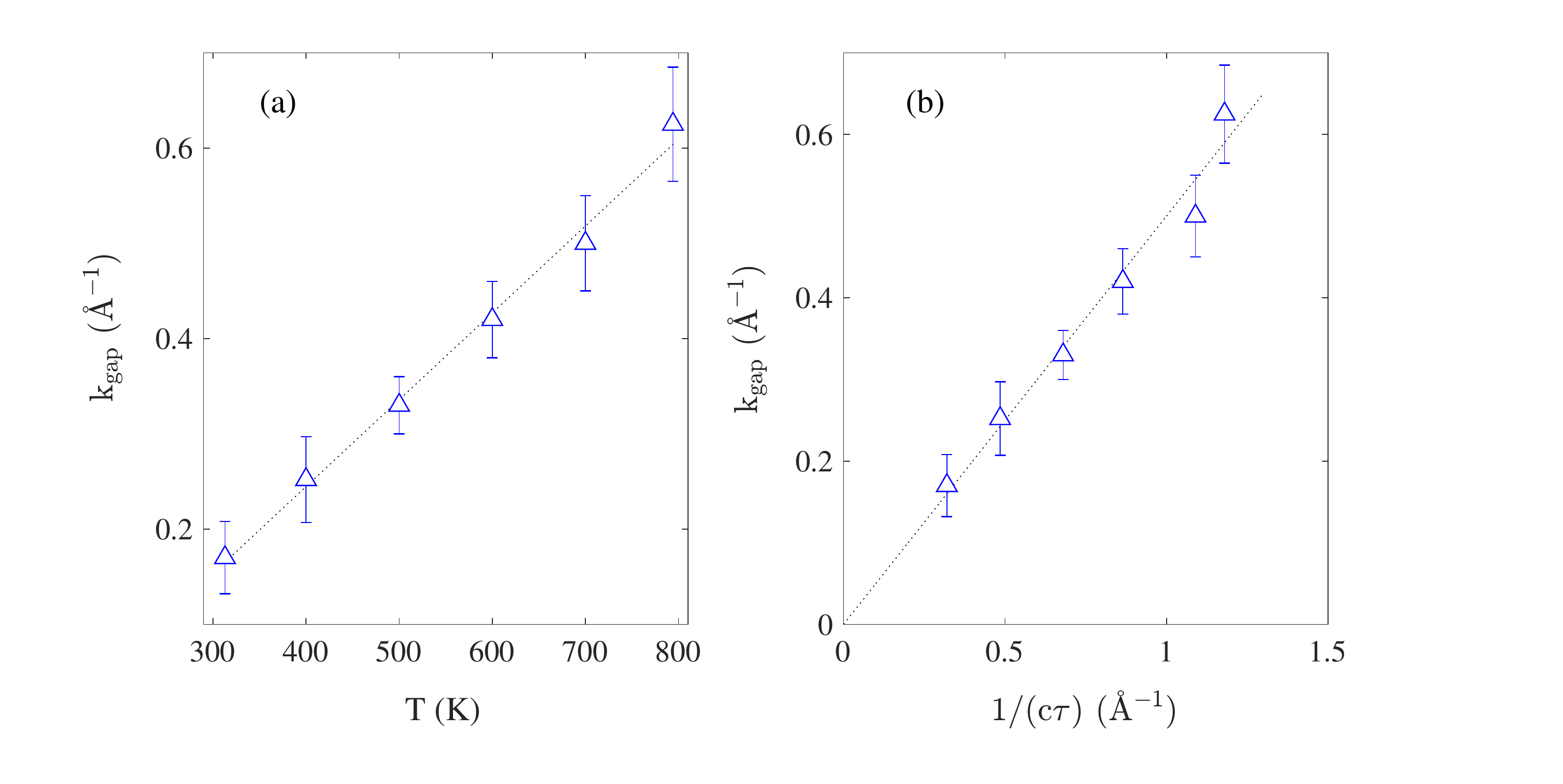} %[height=7.5cm, angle=0]{fig05.eps}
			\caption{(a): Temperature dependence of the width of ``gap'' in the transverse dispersion law. The dashed line represents the linear fit $k_{gap}=\alpha\cdot T+k_0$ with the coefficients $\alpha=9.12\cdot10^{-4}$~K$^{-1}$\rm{\AA}$^{-1}$, $k_0=-0.12$~\rm{\AA}$^{-1}$. (b): Relationship between the gap width and the characteristic quantity $1/c \tau$. The dashed line represents the linear fit $k_{gap}=0.5/c \tau$.}
			\label{Fig_Speed}
		\end{center}
	\end{figure*}

	Second, we plot the $k$ gap as a function of temperature in Fig. \ref{Fig_Speed} and observe its increase. This is consistent with the prediction of Eq. (\ref{kg}) because $\tau$ decreases with temperature.
	
	Third and finally, Eq. (\ref{kg}) predicts the increase of $k_g$ as $1/c\tau$. In Fig. \ref{Fig_Speed}, we plot $k_g$ as a function of $1/c\tau$ and observe a linear dependence, consistent with the theoretical prediction.
	
	\section{Discussion and summary}
	
	This joint experimental, computational, and theoretical study represents further progress towards an understanding of collective modes in liquids and GMS in particular. With new INS data, the longitudinal and transverse collective modes predicted by the EAM potential for gallium enjoys agreement with both IXS and INS experiments. The longitudinal modes additionally sport very solid agreement from low $k$ up to and beyond the first BZ boundary. This increases our confidence that experiments, theory and modelling have now developed to the extent where they reliably describe the same physical mechanism. In the area of liquids, achieving this has remained a long-standing challenge form the point of view of experiment, theory and modelling \cite{Trachenko2016}.
	
	Our current evidence for GMS comes from theory and modelling which agree well, as follows from our Figs. \ref{Fig_TransDisp} and \ref{Fig_Speed}. Moreover, theory and modelling also agree with INS and IXS experiments at larger $k$. Although we do not directly observe GMS in the current INS experiment (due persisting challenges of detecting a weaker transverse mode in liquids at low $k$ \cite{Trachenko2016}), this agreement, together with the agreement of all three lines of enquiry for the longitudinal mode, builds up the body of evidence for GMS. Our results serve as a stimulus for future INS experiments investigating transverse modes in liquids at low $k$ and their evolution in terms of gapped momentum states.
	
	We find that the theoretical prediction of GMS following from Maxwell-Frenkel theory describes liquid dynamics with a fairly high degree of accuracy. This is important for understanding most basic dynamical and thermodynamic properties of liquids as discussed in the Introduction. The overarching goal of this research programme is to reach the stage where, despite the complexity of their theoretical description \cite{landau}, liquids emerge as systems amenable to theoretical understanding at the level comparable to gases and solids.
	
	We are grateful to EPSRC (Grant No. EP/R004870/1) and Russian Science Foundation (project No. 19-12-00022). The molecular dynamic simulations were performed by using the computational cluster of Kazan Federal University and the computational facilities of Joint Supercomputer Center of Russian Academy of Sciences.
	Experiments at the ISIS Neutron and Muon Source were supported by a beamtime allocation RB1820566 from the Science and Technology Facilities Council.

\end{document}